%% file: main.tex
\documentclass[sigconf,nonacm]{acmart}

\AtBeginDocument{%
  \providecommand\BibTeX{{%
    \normalfont B\kern-0.5em{\scshape i\kern-0.25em b}\kern-0.8em\TeX}}}

\begin{document}

\title{Deploying in-network caches in support of distributed scientific data sharing}

\author{Alex Sim}
%\email{webmaster@marysville-ohio.com}
\affiliation{%
  \institution{Lawrence Berkeley National Laboratory}
%  \streetaddress{P.O. Box 1212}
  \city{Berkeley}
  \state{California}
  \country{USA}
}
\email{asim@lbl.gov}

\author{Ezra Kissel and Chin Guok}
\affiliation{%
  \institution{Energy Sciences Network}
  \city{Berkeley}
  \state{California}
  \country{USA}
}
\email{{kissel, chin}@es.net}

\renewcommand{\shortauthors}{Sim and Kissel, et al.}

\begin{abstract}
The importance of intelligent data placement, management, and analysis has become apparent as scientific data volumes across the network continue to increase. To that end, we describe the use of in-network caching service deployments as a means to improve application performance and preserve available network bandwidth in a high energy physics data distribution environment. Details of the software and hardware deployments, performance considerations, and cache usage analysis will be described. We include thoughts on possible future deployment models involving caching node installations at the edge along with methods to scale our approach.
\end{abstract}

%\snowmass

%\begin{CCSXML}
%\end{CCSXML}

%\ccsdesc[500]{Computer systems organization~Embedded systems}

%\keywords{datasets, neural networks, gaze detection, text tagging}


\maketitle

\input{introduction}
\input{background}
\input{analysis}

\input{deployment}
%\input{related}
%\input{future}
\input{conclusion}

\begin{acks}
This work was supported by the Office of Advanced Scientific Computing Research, Office of Science, of the U.S. Department of Energy under Contract No. DE-AC02-05CH11231, and also used resources of the National Energy Research Scientific Computing Center (NERSC). 
\end{acks}

\bibliographystyle{ACM-Reference-Format}
\bibliography{references}

%\appendix



\end{document}

%% file: introduction.tex
\section{Introduction}
\label{sec:introduction}
With advances in instruments and computing hardwares, scientific experiments and simulations generate an increasing amount of data, and share the data among geographically distributed users. 
Existing datasets and the cost of the storage hardware and its maintenance limit the number of data sources, and data distribution requires higher network bandwidth for timely data delivery. 
Some datasets are popular among the users and transferred multiple times to the same institution or institutions in the same region for the different users as well as for the same user for various reasons. Also, users move the data from the data sources to multiple computing locations depending on when the computing jobs run for the user.
Some type of content delivery network or hierarchical data distribution infrastructure can accommodate sharing data among  geographically distributed users by pre-staging popular datasets that many users might use in the same region or same institution. 

In-network caching provides a different type of content delivery network for scientific data infrastructure, supporting on-demand temporary caching service. It also enables for a network provider to design data hotspots into the network topology, and to manage traffic movement and congestion by data-driven traffic engineering.
Regional in-network caching strategy would also reduce the data access latency for the users and increase the overall computing application performance.
For the network providers, in-network caching service would decrease traffic bandwidth demands on busy links. This is especially relevant to the High Energy Physics (HEP) community with the LHC instrument at CERN and Tier-1 sites for the ATLAS at Brookhaven National Laboratory and CMS experiments at Fermi National Accelerator Laboratory.

For example, Southern California Petabyte Scale Cache (SoCal Repo)~\cite{socalrepo2018} based on XCache~\cite{xrootdcms,xcache2014,stashcache,Fajardo2020} consists of 24 data cache nodes with approximately 2.5PB of storage space, supporting client computing jobs for High-Luminosity Large Hadron Collider (HL-LHC) analysis in Southern California. The SoCal Repo has cache nodes at the ESnet junction in Sunnyvale, at Caltech, and at UCSD. The system has been used by the CMS collaboration for real CMS data analysis as part of the Caltech and UCSD Tier-2 center production infrastructure. We observed that by sharing dataset in the regional data cache, the network traffic demand was reduced by a factor of 2 on the average over the observed period. Studying the characteristics of the data access patterns~\cite{copps2021} has enabled new strategies for how the needed resources such as compute, storage, and network can be allocated.

In this paper, we build upon our prior analysis work ~\cite{copps2021} by providing an update on the existing caching infrastructure, and we include sample analysis results from HEP jobs using the SoCal Repo since our most recent node deployments have not yet produced any data.
We introduce DTN-as-a-Service (DTNasS) as our approach for hosting in-network caching and storage services for scientific datasets within ESnet, and we describe our methodology and experiences in deploying new caching hardware within ESnet points-of-presence. Finally, we discuss future deployment opportunities driven by our data-driven analysis and flexible provisioning framework.

%% file: background.tex
\section{Background}
\label{sec:background}

\subsection{Energy Sciences Network (ESnet)}
The Energy Sciences Network (ESnet) is the US Dept of Energy (DOE) Office of Science’s high-performance network user facility, delivering highly-reliable data transport capabilities optimized for the requirements of large-scale science. ESnet is stewarded by the Advanced Scientific Computing Research Program (ASCR) and managed and operated by the Scientific Networking Division at Lawrence Berkeley National Laboratory (LBNL).  ESnet acts as the primary data circulatory system for science by interconnecting the DOE’s national laboratory system, dozens of other DOE sites,
and 150 research and commercial networks around the world. It allows tens of thousands of scientists at DOE laboratories and academic institutions across the country to transfer vast data streams and access remote research resources in real-time. ESnet exists to provide the specialized networking infrastructure and services required by the national laboratories, large science collaborations, DOE user facilities, and the DOE research community. All together, ESnet provides a foundation for the nation’s scientists to collaborate on some of the world’s most important scientific challenges, including energy, biosciences, materials, and the origins of the universe. Science data traffic across its network has grown
at around 60\% each year, and traffic has exceeded an exabyte per year since 2019.

\subsection{High Energy Physics (HEP)}
The HEP experiments have been generating large volume of data~\cite{esnetHepReq}, especially from the LHC in Switzerland. Experiments such as ATLAS and CMS have thousands of globally geographically distributed collaborations, and the efficient data distribution infrastructure has been explored for a long time, delivering Petabytes of data. The LHC community has been preparing the increase in annual data volume, and one of the approaches is the data replication between regional "Data Lakes"~\cite{datalakes} and the mixture of remote access and caching within those lakes. A regional data lake is expected to serve multiple computing centers within that region.  

%\subsection{Open Science Grid (OSG)}

\subsection{DTN-as-a-Service (DTNaaS)}

Both the Open Science Grid (OSG) \cite{osg} and broader HEP community have embraced software containers as part of their data and analysis pipelines where container deployment solutions such as HELM+Kubernetes \cite{helm} and SLATE-CI \cite{slate} have become commonplace. As a result, a number of existing service container workflows (often using Docker and Docker Hub) are being maintained in public repositories and container registries. Open source software and tooling around containers is also feature-rich and improving with time. In contrast with virtual machines (VMs), containers are generally a lighter-weight option that support the execution environment for single services, and have benefits when ease of software packaging, distribution, and upgrade cycles are considered. %Building and supporting versioned VM images is often a less flexible approach compared to containers in terms of rapid deployment, ease of configuration, and %when considering the ease of maintaining packaged software release dependencies over time.

The deployment we describe makes use of a platform known as DTN-as-a-Service (DTNaaS) \cite{janus} being developed within ESnet, which provides a tailored container orchestration framework focused on the configuration and tuning options relevant to high-performance data movement services. Underpinning the design philosophy was a recognition that more general, feature-rich platforms such as Kubernetes may not always provide an ideal interface for single-service container instances where automated resource scaling and migration are not primary concerns. The ability to control specific features at the container level, such as attaching multiple network interfaces (and interface types), and dual-stack configurations for IPv4 and IPv6 at a fine granularity with minimal setup cost, were primary design goals. While the initial focus has been on traditional Data Transfer Node (DTN) software endpoints, the framework has been designed to be flexible enough to accommodate many types of software containers that have high-throughput and advanced networking configuration requirements. Thus, a DTNaaS solution that can support performance-oriented data mover services with modest support overheads was an ideal fit for deploying OSG XCache software containers from an "in-the-network" perspective.

%% file: analysis.tex
\section{Cache Usage Analysis}
\label{sec:analysis}

We collected data access measurements from the SoCal Repo between July 2021 and Dec. 2021, where HL-LHC analysis jobs requested data files for users. The SoCal Repo has 24 cache nodes regionally at Caltech, UCSD and ESnet, consisting of approximately 2.5PB of storage. We studied how much data is shared, how much network traffic volume is consequently saved, and how much the in-network data cache contributes to the resource management and performance. Additionally, we analyzed data access patterns and observed the impacts of new cache nodes to the regional data repository. The data access pattern study may show a few characteristics of the data sharing, cache utilization, and network utilization where shared data directly contributes to the network traffic savings.

\begin{table}[h!]
\scriptsize
\centering
\caption{
Summary statistics for data accesses at the SoCal Repo from July to Dec. 2021
}
\begin{tabular}{|c||c|c|c|} \hline
  & \# of accesses & data transfer size (TB) & shared data size (TB) \tabularnewline \hline \hline
July 2021 & 1,182,717 & 385.78 & 519.25 \tabularnewline \hline
Aug 2021 & 1,078,340 & 206.94 & 313.46 \tabularnewline \hline
Sep 2021 & 1,089,292 & 206.96 & 257.18 \tabularnewline \hline
Oct 2021 & 1,058,071 & 412.18 & 141.91 \tabularnewline \hline
Nov 2021 & 878,703 & 649.30 & 82.67 \tabularnewline \hline
Dec 2021 & 983,723 & 1,257.89 & 130.03 \tabularnewline \hline
Total &  6,270,846 & 3,119.07 & 1,444.51 \tabularnewline \hline
Daily average & 34,838.03 & 17.42 & 8.03 \tabularnewline \hline
\end{tabular}
\label{tab:summary_data_all}
\end{table}

Table \ref{tab:summary_data_all} shows the basic statistics on the data access activities for all caching nodes during the study period (from July to Dec. 2021). 
The "data transfer size" column in Table \ref{tab:summary_data_all} indicates the data volume resulting from the "cache misses", when a data file was transferred from the remote data source. For cache misses, any caching nodes in the SoCal Repo did not have the data, resulting in a data transfer from the remote site to one of the caching nodes. 
The "shared data access size" column refers to the data volume from the "cache hits", when the data file was shared from the cache. The shared data accesses correspond to the network traffic savings.
As the new cache nodes were added every month since Sep 2021, the percentage of shared data size decreases with time because the requests would fill the new cache nodes first by the policy. The  percentage of the share size would be maintained at a certain level after the cache nodes are full.

\begin{figure}[htb!]
\centering
\includegraphics[width=\columnwidth]{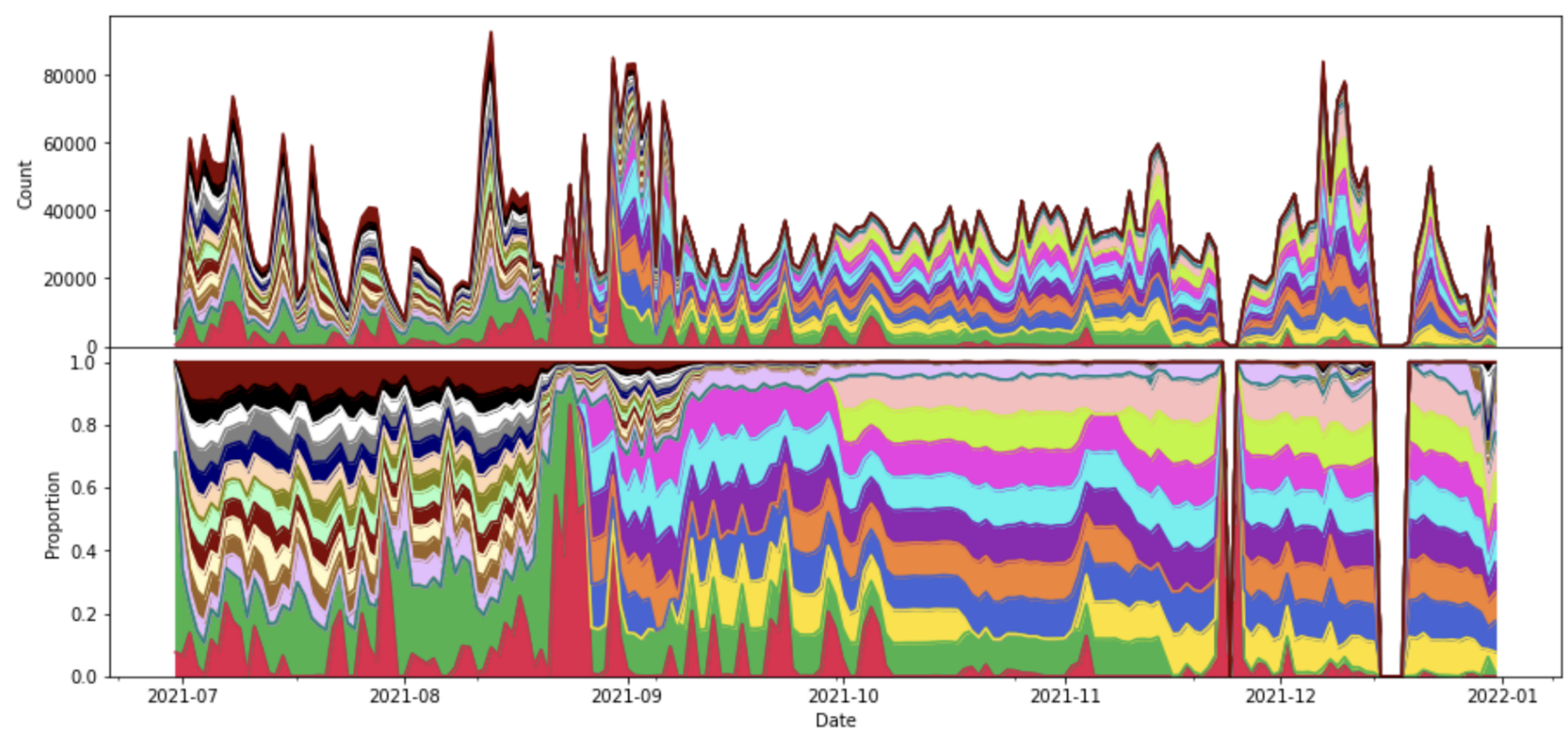}
\caption{
    Daily total data access sizes and proportion of total access sizes on each node in SoCal Repo
}
\label{fig_access_all}
\end{figure}

Figure \ref{fig_access_all} shows the daily total data access size (in TB) to the SoCal Repo including both cache hit sizes and cache miss sizes, and ratio among the cache nodes. It indicates that there are new nodes added to the SoCal Repo since Sep. 2021 showing that the majority of the activities, in particular data transfers occurred on new cache nodes since Sep. 2021.  

\begin{figure}[htb!]
\centering
\includegraphics[width=\columnwidth]{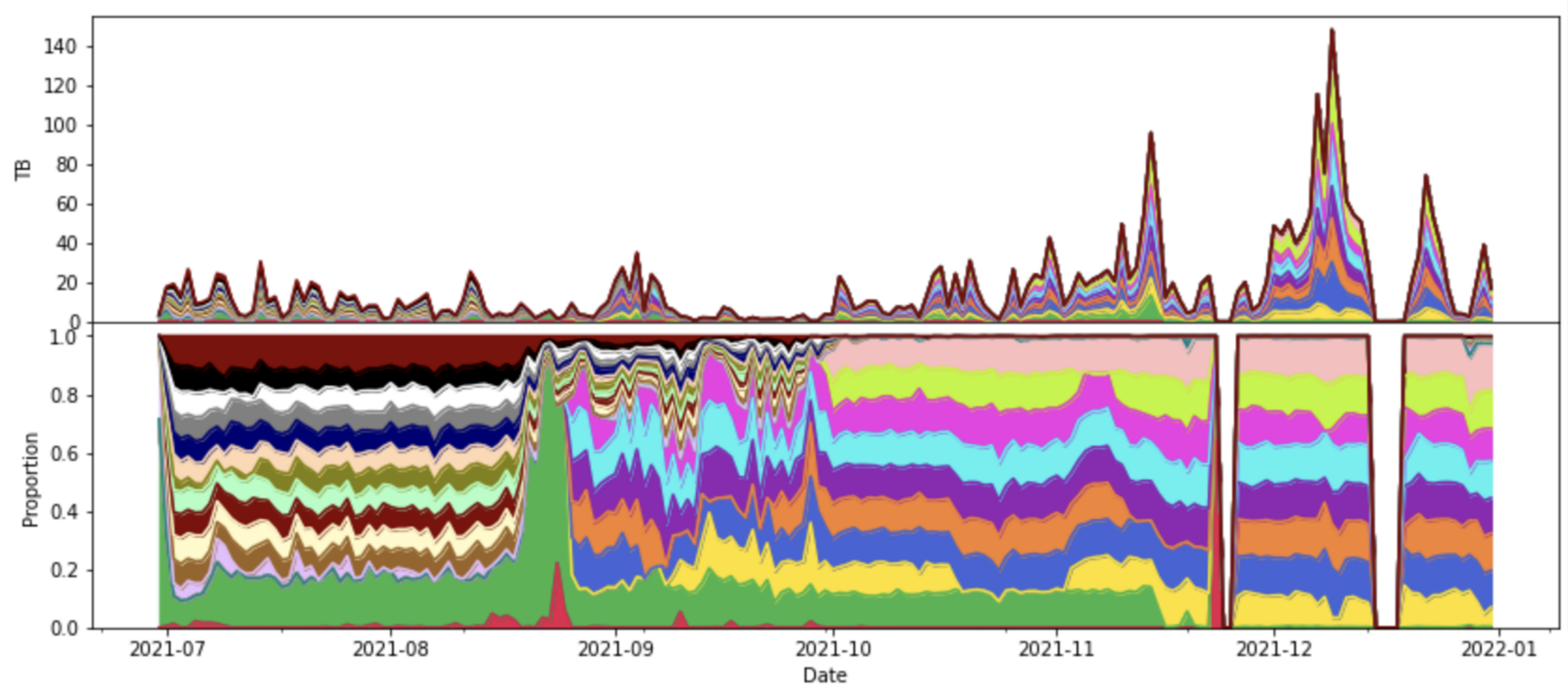}
\caption{
    Daily total cache miss sizes (data transfer volume) and proportion of each node in SoCal Repo
}
\label{fig_cache_misses}
\end{figure}

\begin{figure}[htb!]
\centering
\includegraphics[width=\columnwidth]{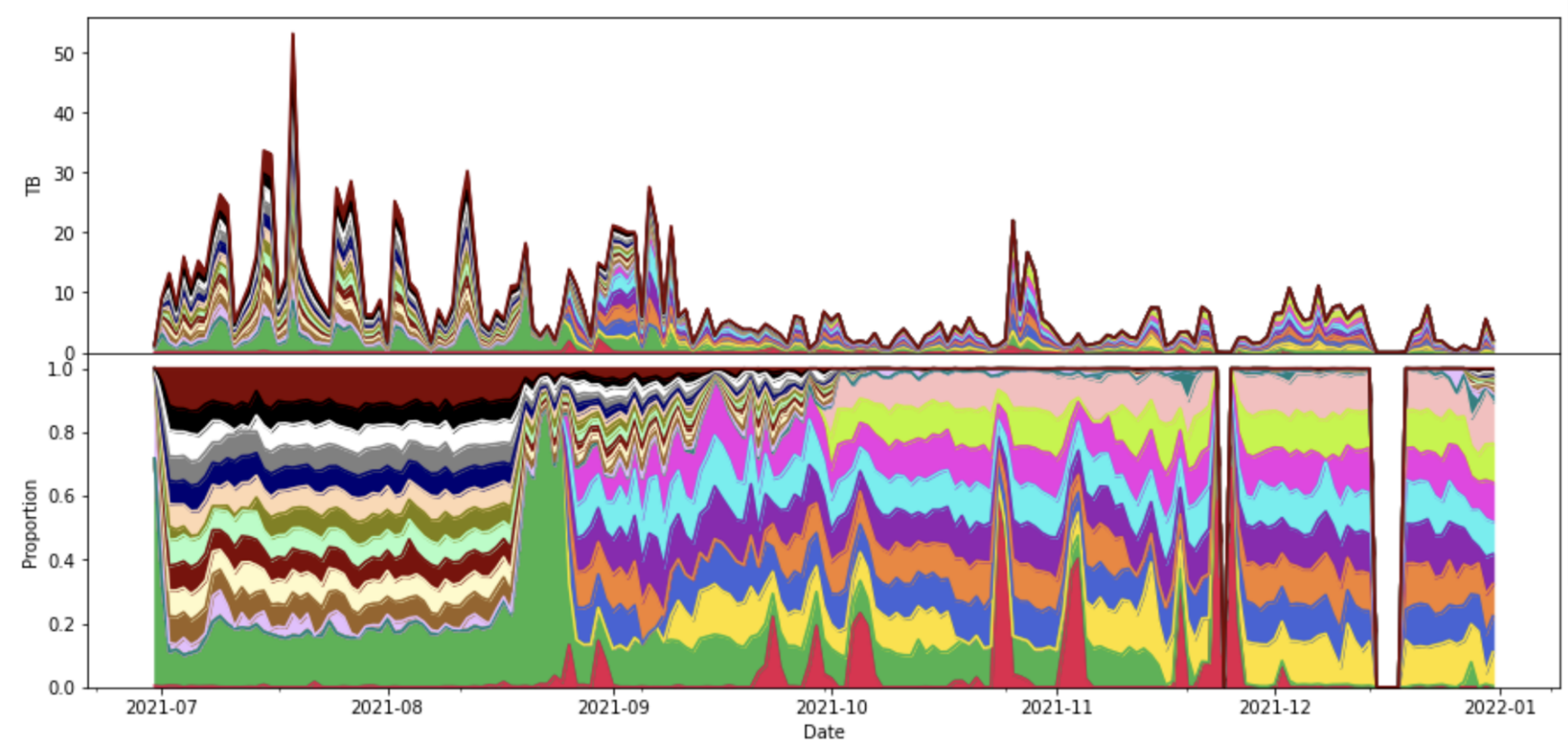}
\caption{
    Daily total cache hit sizes (shared data volume) and proportion of each node in SoCal Repo
}
\label{fig_cache_hits}
\end{figure}

Figure \ref{fig_cache_misses} shows the daily data transfer sizes (in TB) from remote sites to the SoCal Repo upon the user request for the data. 
Figure \ref{fig_cache_hits} shows the daily shared data sizes (in TB) which corresponds to the network traffic savings by the repeated accesses to the same data (cache hits).
Both Figures \ref{fig_cache_misses} and \ref{fig_cache_hits} indicates a new trend since the new cache nodes, which is 10 times larger in the storage space than the existing nodes, have been added to the SoCal Repo in Sep. 2021, and more data transfer activities are observed on new cache nodes.

\begin{figure}[htb!]
\centering
\includegraphics[width=\columnwidth]{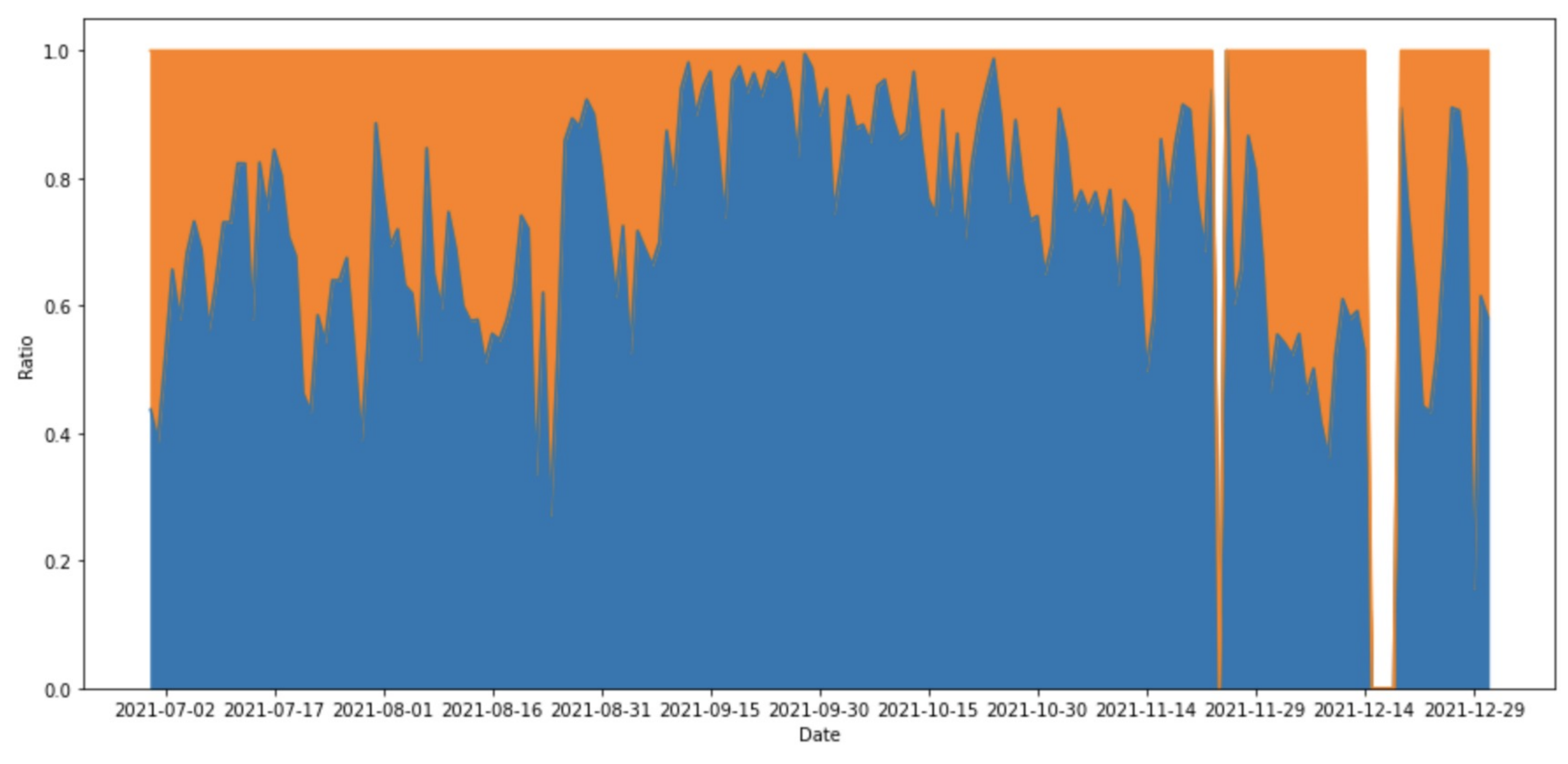}
\caption{
    Daily proportion of the cache misses and cache hits in SoCal Repo
}
\label{fig_cache_prop}
\end{figure}

Figure \ref{fig_cache_prop} shows the daily proportion of the cache misses and cache hits, and also shows that there are a large portion of the cache hits in the daily accesses indicating network traffic savings. 

\begin{figure}[htb!]
\centering
\includegraphics[width=\columnwidth]{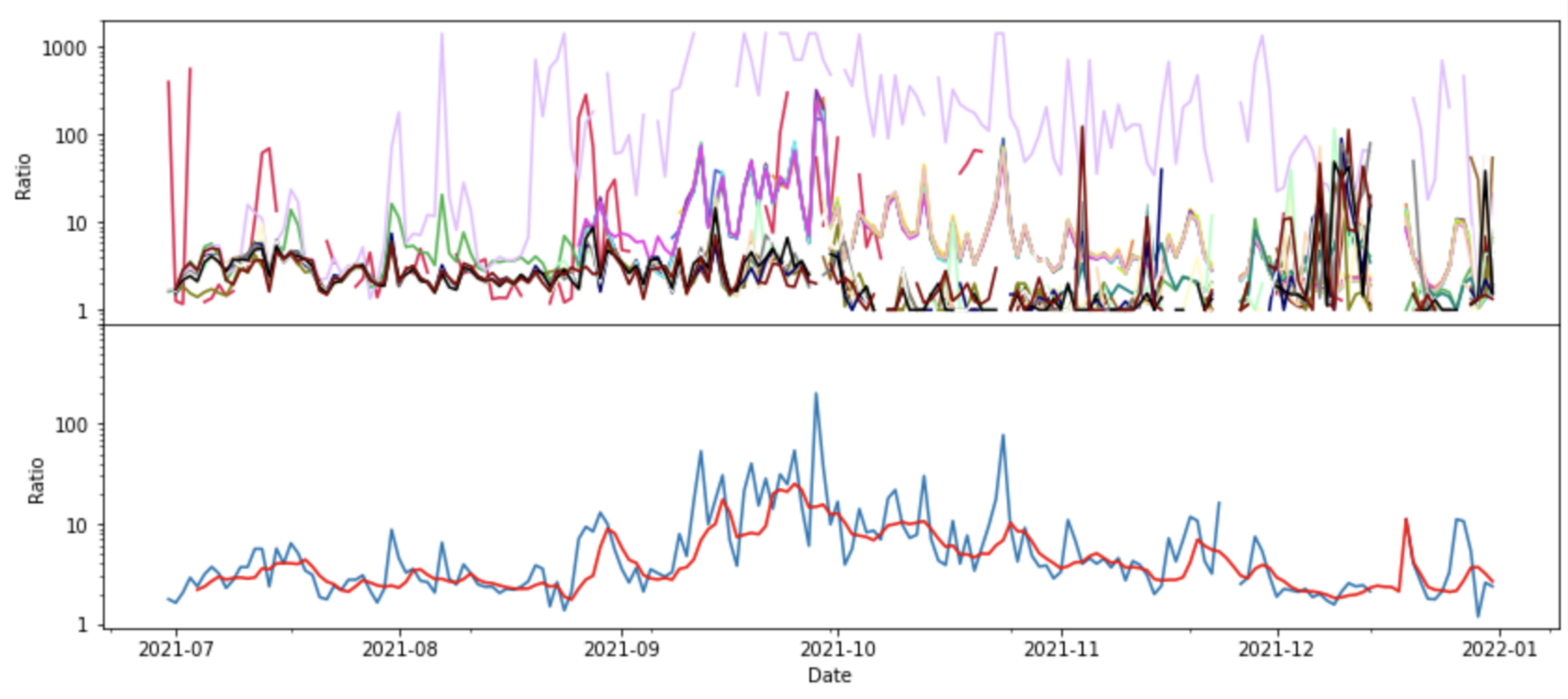}
\caption{
    Daily network traffic frequency reduction rates in SoCal Repo
}
\label{fig_net_freq}
\end{figure}

Figure \ref{fig_net_freq} shows the daily traffic frequency reduction rates from the data sources to the local cache, indicating the ratio of the number of daily data transfers occurring regardless of the data transfer size. The average traffic frequency reduction rate is 3.43 over the study period. Some nodes indicate very high rates as the most of the connections are the shared data in the local cache. 

\begin{figure}[htb!]
\centering
\includegraphics[width=\columnwidth]{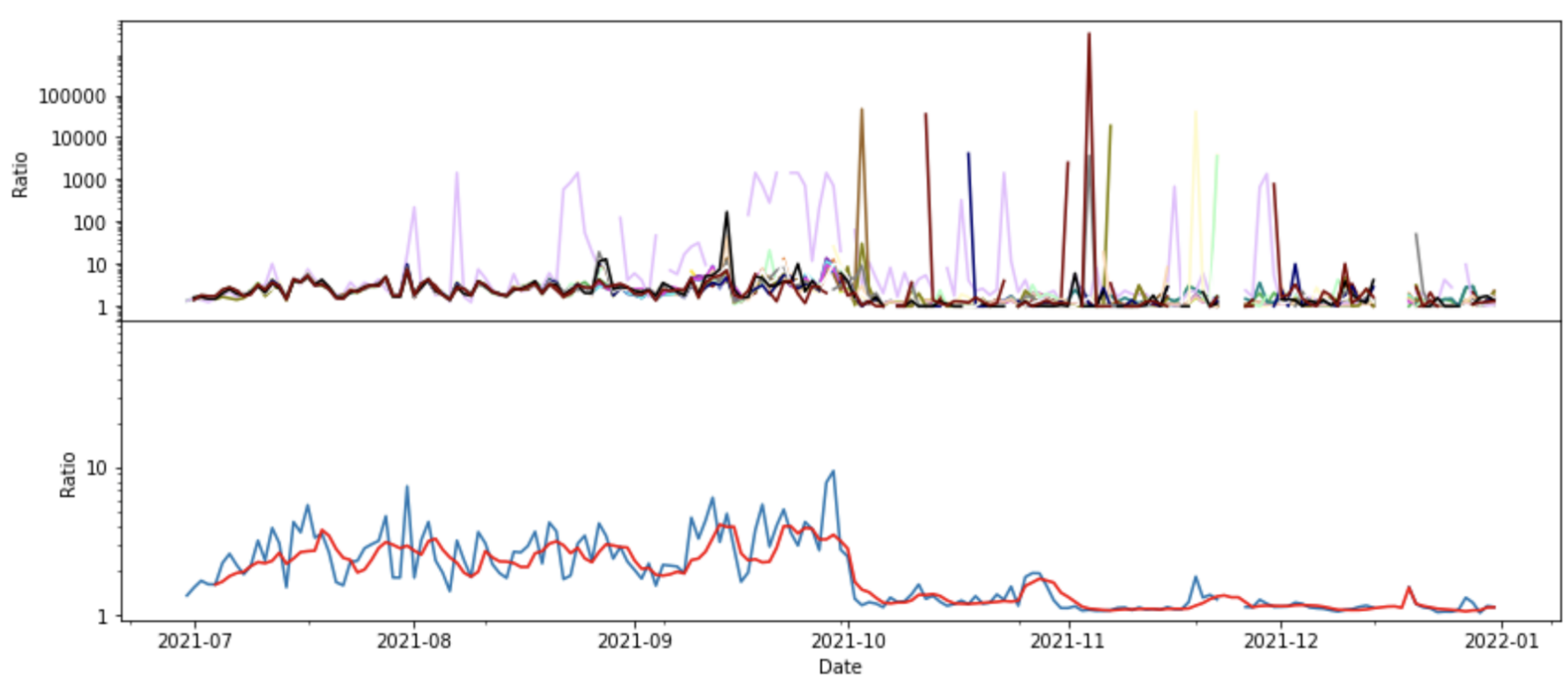}
\caption{
    Daily network traffic volume reduction rates in SoCal Repo
}
\label{fig_net_vol}
\end{figure}

Figure \ref{fig_net_vol} shows the daily network traffic volume reduction rates, indicating the ratio of the daily total data volume being moved from remote sites to the local cache. It also shows the effect of the new cache nodes being added to the SoCal Repo since Sep. 2021 while the existing cache nodes maintain higher rates of the shared data. The red line on the bottom plot is for the moving average over a week.
The average traffic volume reduction rate is 1.47 during the study period (1.68 until Nov 2021).

\begin{figure}[htb!]
\centering
\includegraphics[width=\columnwidth]{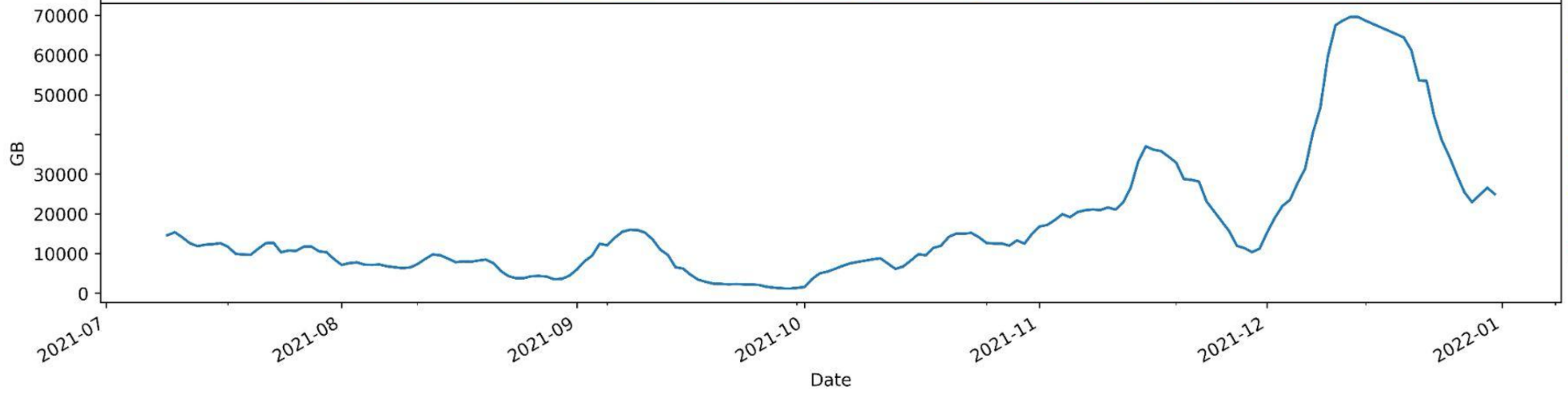}
\caption{
    Cache miss size with 1-week moving average in SoCal Repo
}
\label{fig_cache_miss_1wkma}
\end{figure}

\begin{figure}[htb!]
\centering
\includegraphics[width=\columnwidth]{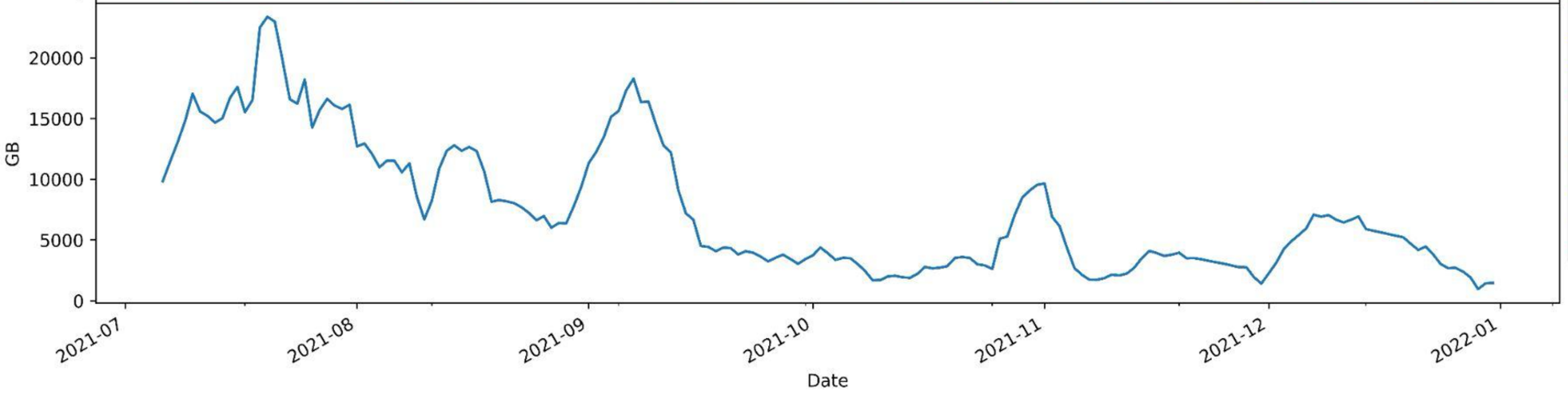}
\caption{
    Cache hit size with 1-week moving average in SoCal Repo
}
\label{fig_cache_hit_1wkma}
\end{figure}

Figure \ref{fig_cache_miss_1wkma} shows the data transfer sizes for cache misses with 1-week moving average. 
Figure \ref{fig_cache_hit_1wkma} shows the shared data sizes for cache hits with 1-week moving average.

%% file: deployment.tex
\section{Cache Node Deployments}
\label{sec:deployment}

\begin{figure}
    \centering
    \includegraphics[width=0.48\textwidth]{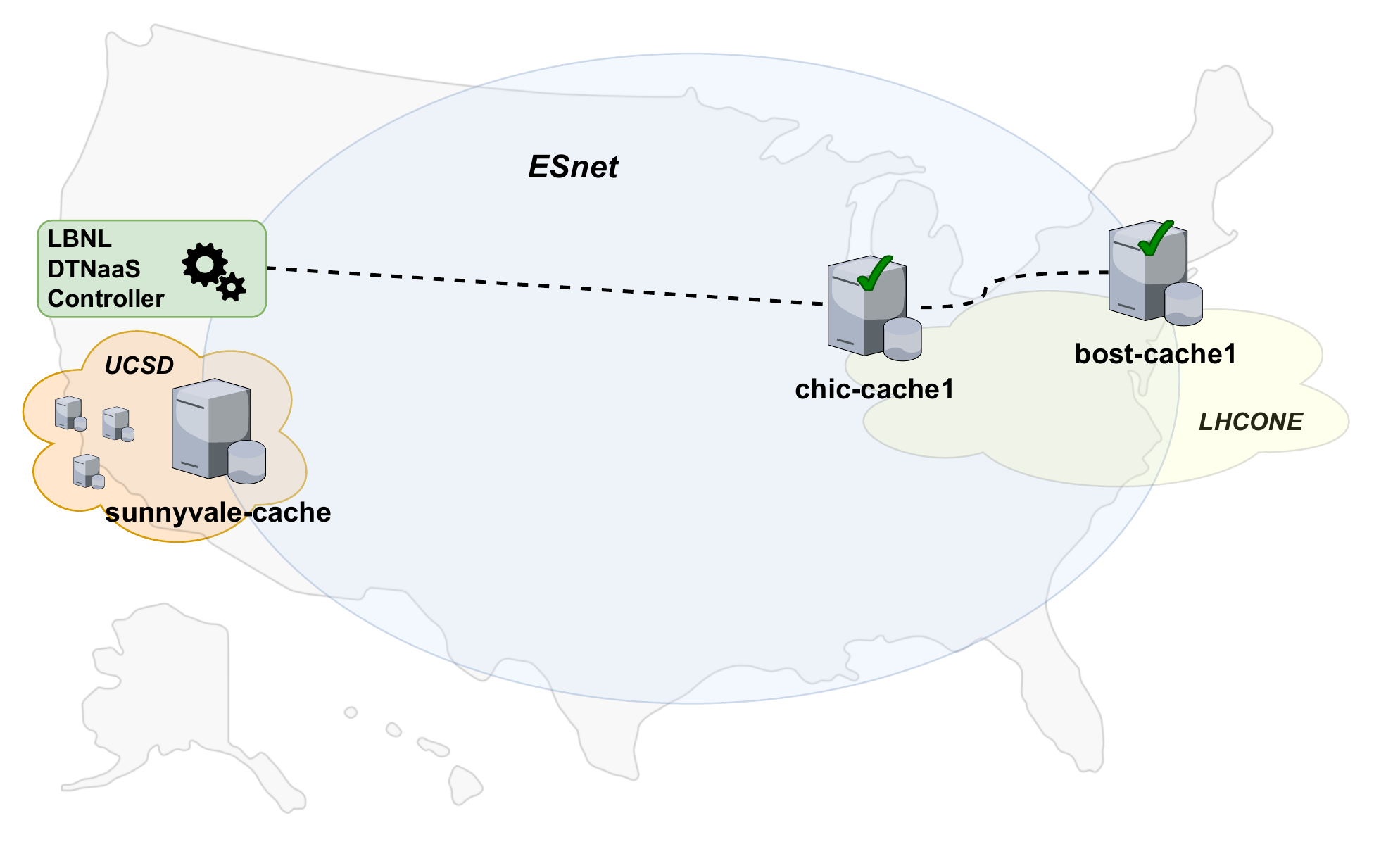}
    \caption{US deployment of caches within ESnet. Two new caching node deployed in Boston and Chicago with container services managed by a DTNaaS controller instance located at ESnet's LBNL datacenter.}
    \label{fig:deploy}
\end{figure}
% edit link: https://drive.google.com/file/d/1x_HdRP-0ObGampfHgVsKYywmn_nvncRi/view?usp=sharing

\begin{figure}
    \centering
    \includegraphics[width=0.48\textwidth]{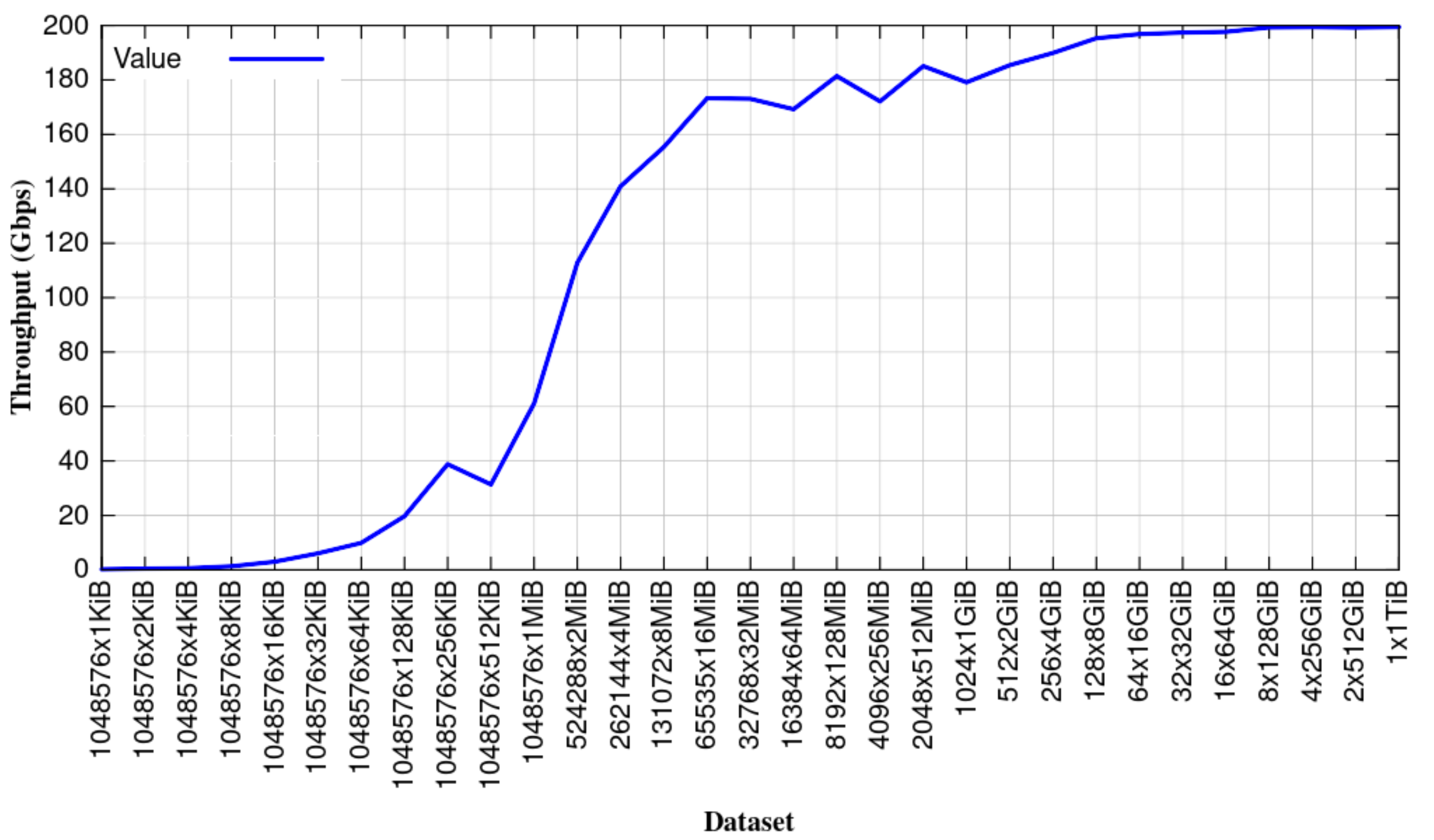}
    \caption{Storage subsystem performance across a range of synthetic data sets.}
    \label{fig:storage-perf}
\end{figure}

Two additional caching nodes have been deployed within ESnet to supplement the existing Sunnyvale node within the broader SoCal Repo (Fig. \ref{fig:deploy}). These nodes were installed as physical servers at ESnet points of presence (PoP) in Boston and Chicago. The specifications of each node is as follows: dual-socket Intel Xeon Gold 5220S CPUs, 384GB RAM, 12x Micron 9300 PRO 15.36TB NVMe SSDs, and Mellanox ConnectX-5 100G network interface cards (NICs). The node storage was configured to provide an effective 165TB of available capacity each while providing suitable performance to match the expected XCache usage given 100G network connectivity. Figure \ref{fig:storage-perf} shows the results of using the elbencho\footnote{elbencho: \url{https://github.com/breuner/elbencho}} storage benchmarking tool on the caching filesystem exposed to a container.

Part of our strategy has been to explore the logistics of hosting data movement services within an international science network such as ESnet where \emph{networking} services, not \emph{application} services, have been the traditional offering. The development of DTNaaS is one ongoing effort to help bridge this technology gap, and below we describe some of the architectural and technical decisions that have been put into practice.

\subsection{Network Connectivity}

\begin{figure}
    \centering
    \includegraphics[width=0.45\textwidth]{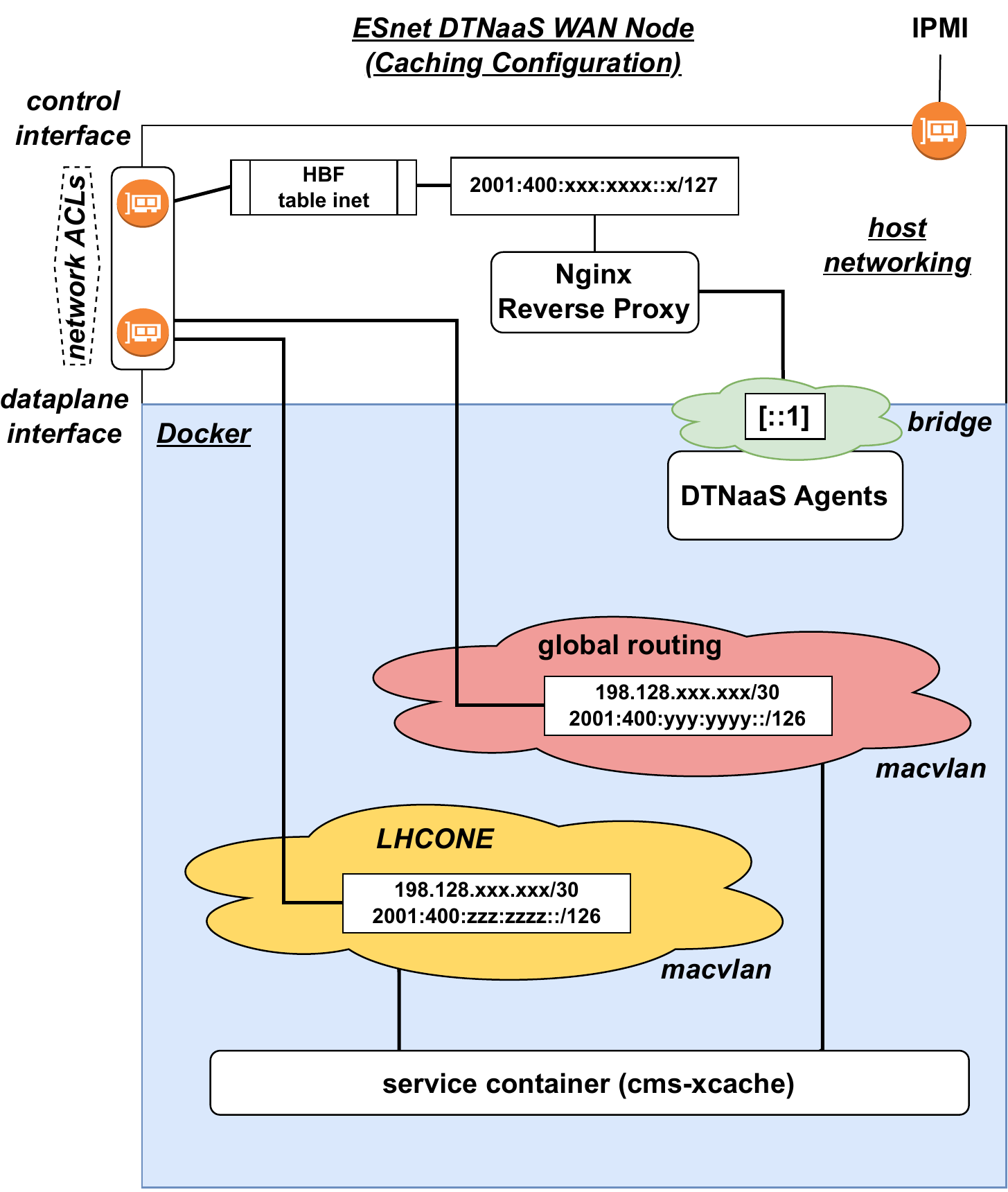}
    \caption{The DTNaaS networking configuration for cms-xcache deployments. Each service container is dual-homed and dual-stacked (IPv4, IPv6) using \emph{macvlan} interfaces for the high-speed dataplane. The control interface provides infrastructure and service management connectivity.}
    \label{fig:dtnaas-net}
\end{figure}

A typical DTNaaS deployment relies on a centralized controller that contacts one or more nodes to provision and manage service containers. The controller runs in a data center environment while the DTNaaS service nodes are located at a wide-area network point-of-presence. The low-bandwidth communication between the controller and service nodes forms the management network and is achieved using a secured, dedicated control interface as diagrammed in Figure \ref{fig:dtnaas-net}. A number of agent processes are run in containers to provide the DTNaaS management functionality over this control connection, and their access is facilitated by using a standard Docker bridge network with processes binding to localhost ([::1]). A reverse proxy provides encrypted external endpoints to these locally bound processes via the host networking namespace.

The high-speed dataplane is realized through physical connections from the node's Mellanox NIC ports directly to ESnet routers, and flows originating from or terminating at the XCache service container make use of these links.  Key requirements for these connections included effective network isolation from the baremetal host OS and low virtualization overhead to achieve line rate performance. To that end, \emph{macvlan} container networks were configured in 802.1q trunk bridge mode and attached to the parent VLAN-tagged service interfaces instantiated in the host namespace. The DTNaaS controller was then able to map the dataplane networks to the XCache container configuration and automatically expose the \emph{macvlan} interfaces to each running container as appropriate. 

Generally, the Linux \emph{macvlan} sub-interface type provides isolation from the host networking namespace at layer 2. From an external ESnet switch or router perspective, a container with a macvlan interface will appear as another unique host on the network with a unique MAC and/or IP address. For the baremetal host to communicate with the running container, another \emph{macvlan} sub-interface must be created on the host and local routing adjusted for any configured layer 3 subnet(s). In a similar fashion, another container may also communicate within the \emph{macvlan} subnet if attached to the same brdiged container network. In this manner DTNaaS service containers provide separation between networks allocated for each service container and reduce the complexity of a single filtering rule set maintained alongside the host network namespace.

An additional complexity in our deployment involves the requirement that our XCache containers are running both in a dual-home and dual-stack configuration. As shown in Figure \ref{fig:dtnaas-net}, two separate routing instances are exposed to the containers via \emph{macvlan} interfaces: 1) a global routing service to provide a default route and support common networking tasks such as name resolution, and 2) an LHCONE L3VPN service to provide connectivity to other HEP data sources and peers. In each instance, both IPv4 and IPv6 addressing is configured in the container networks. To support this connectivity model, the DTNaaS controller was extended so that multiple networks (i.e. interfaces) could be mapped to service profiles and dual-stack addressing specified for each network instance, allowing our multi-node deployment to be managed in a centralized and automated manner.

\subsection{Network Security}

The DTNaaS management network is secured using a combination of a reverse proxy service (as noted above) and host-based filtering using standard \emph{iptables} rules. In addition, our deployment makes use of network ACLs that are implemented on the ESnet router platform to provide ingress and egress filtering for both the control and dataplane interfaces on each node. The routing instances exposed to the XCache container via the \emph{macvlan} networks have unique ACLs applied; for example, only the XCache TCP port is allowed for ingress in the LHCONE instance. This ability to rely on the network infrastructure to provide packet filtering simplifies any otherwise necessary host and container configuration considerably.

In the absence of, or in addition to, network ACL support, \emph{macvlan} bridges can be secured using features available within the Linux kernel's \emph{nftables} subsystem. To implement filtering when using container \emph{macvlan} networks, use of the \emph{netdev} address family table available in \emph{nftables} may be employed. This mechanism allows for inspecting packets as they arrive directly off of a specific interface, before they hit any prerouting chains. Such an ingress hook can be configured from the host network namespace since it accesses the physical interface directly, and this approach affords more flexibility and performance but at the cost of some stateful features. As the Linux netfilter mechanisms continue to evolve \cite{ebpf,ebpf-perf} we expect to integrate additional capabilities into the DTNaaS framework to support a variety of filtering and traffic management options for managed containers.

\subsection{Container Management}

The CMS XCache container images managed by DTNaaS for this deployment are sourced from the OSG Docker Hub Community Organization\footnote{cms-xcache image: https://hub.docker.com/r/opensciencegrid/cms-xcache}. The DTNaaS framework additionally makes use of an internal container registry that performs security vulnerability scanning using Trivy \cite{cont-sec,trivy} as part of a larger constant integration (CI) pipeline for managed images. Automation ensures we pull the latest OSG-maintained images from Docker Hub and have them pass through our CI infrastructure before being instantiated on the physical infrastructure nodes. 

Finally, our DTNaaS approach includes tooling to allow for rapid startup or shutdown of distributed caching instances as well as the ability to maintain a history of image revisions for rollback to a prior version if necessary. A command line interface exposes the most sysadmin-friendly mechanisms for common operational tasks, while a controller API is also available for more programmatic integration.

As of this writing, the installation and verification of the two caching nodes in Boston and Chicago has been completed and we are awaiting final integration into the CMS workflow. The additional caching infrastructure will form the nucleus of Midwestern (Chicago) and Eastern (Boston) regional CMS Data Lakes to improve the data accessibility of nearby computing facilities, and their placement within the ESnet production network is anticipated to aid in delivering data to peering sites with minimal friction. 

%\begin{figure*}[b]
%    \centering
%    \includegraphics[width=\textwidth]{figures/janus_cli.png}
%    \caption[width=\textwidth]{Snapshot of the DTNaaS CLI showing two active XCache services deployed at Boston and Chicago.}
%    \label{fig:cli}
%\end{figure*}

%% file: conclusion.tex
\section{Conclusion and Future Work}
In this paper, we described our design, deployment and experience on DTN-as-a-Service (DTNasS) as our approach for supporting in-network caching and storage services for scientific datasets within ESnet. We also described observations on the data access trends in the existing SoCal Repo for HEP analysis jobs at Caltech and UCSD. 
In-network data cache enables reduction of the redundant data transfers and consequently network traffic savings. During the study period, 1.4PB of the network traffic volume savings was observed in the SoCal Repo, and the study opens other leads to the network engineering and caching policies.
Data access trends such as Figures \ref{fig_cache_hit_1wkma} and \ref{fig_cache_miss_1wkma} would be able to predict the trend of the network traffic demand in the regional cache. 
Also, cache hits analysis such as Figure \ref{fig_cache_hits} would enable different types of the caching policy based on the dataset popularity. We plan to extend this analysis to observe caching trends with the inclusion of the two new nodes as described, and we plan to study data-driven network traffic engineering and traffic volume forecasting as well as locally customized caching policy in the future.

Future directions for our DTNaaS approach include considering the positioning and flexible service deployments in support of science applications over the entirety of the ESnet footprint, where geographic boundary considerations along with data locality requirements will influence placement. The temporal dimension is another consideration for DTNaaS feature development, where on-demand, dynamic provisioning of data services may be suitable for more ephemeral or time-bound acquisition pipelines. We also hope to more generally package this framework as a means for other facilities or end users to effectively manage and optimize containerized data movement services within their own infrastructures.